\documentclass[conference]{IEEEtran}
\IEEEoverridecommandlockouts
\usepackage{cite}
\usepackage{amsmath,amssymb,amsfonts}
\usepackage{algorithm}
\usepackage{algorithmic}
\usepackage{graphicx}
\usepackage{textcomp}
\usepackage{csquotes}
\usepackage{multirow}
\usepackage{enumitem}
\usepackage{xcolor}
\usepackage[top=0.75in, left=0.75in, right=0.75in, bottom=1.1in]{geometry}
\usepackage[bookmarks=false]{hyperref}

\def\BibTeX{{\rm B\kern-.05em{\sc i\kern-.025em b}\kern-.08em
    T\kern-.1667em\lower.7ex\hbox{E}\kern-.125emX}}
\begin{document}

\title{Joint Optimisation of Load Balancing and Energy Efficiency for O-RAN Deployments}

\author{\IEEEauthorblockN{Mohammed M. H. Qazzaz$^{1,2}$\qquad Abdelaziz Salama$^{1}$ \qquad Maryam Hafeez$^{1}$ \qquad  Syed A. R. Zaidi$^{1}$}
\IEEEauthorblockA{\textit{$^{1}$ School of Electronic and Electrical Engineering,}
\textit{University of Leeds,} Leeds, UK \\
\textit{$^{2}$ College of Electronics Engineering,}
\textit{Ninevah University,} Mosul, Iraq} 
\{M.M.H.Hameed, A.M.Salama, M.Hafeez, S.A.Zaidi\}@leeds.ac.uk}


\maketitle

\IEEEpubidadjcol

\begin{abstract}

Open Radio Access Network (O-RAN) architecture provides an intrinsic capability to exploit key performance monitoring (KPM) within Radio Intelligence Controller (RIC) to derive network optimisation through xApps. These xApps can leverage KPM knowledge to dynamically switch on/off the associated RUs where such a function is supported over the E2 interface. Several existing studies employ artificial intelligence (AI)/Machine Learning (ML) based approaches to realise such dynamic sleeping for increased energy efficiency (EE). Nevertheless, most of these approaches rely upon offloading user equipment (UE) to carve out a sleeping opportunity. Such an approach inherently creates load imbalance across the network. Such load imbalance may impact the throughput performance of offloaded UEs as they might be allocated a lower number of physical resource blocks (PRBs). Maintaining the same PRB allocation while addressing the EE at the network level is a challenging task. To that end, in this article, we present a comprehensive ML-based framework for joint optimisation of load balancing and EE for ORAN deployments. We formulate the problem as a multi-class classification system that predictively evaluates potential RU configurations before optimising the EE, mapping network conditions to three load balance categories (Well Balanced, Moderately Balanced, Imbalanced). Our multi-threshold approach (Conservative, Moderate, Aggressive) accommodates different operational priorities between energy savings and performance assurance. Experimental evaluation using 4.26 million real network measurements from simulations demonstrates that our Random Forest model achieves 98.3\% F1-macro performance, representing 195\% improvement over traditional baseline strategies. The framework enables up to 11.6\% quality of service (QoS) improvement and 82\% better load distribution while maintaining energy optimisation objectives. Cross-scenario validation across different network scales confirms robust generalisation capabilities, making this approach suitable for practical O-RAN deployments requiring intelligent energy management without performance degradation.

\end{abstract}


\section{Introduction}\label{introduction}

Open Radio Access Network (O-RAN) architecture fundamentally transforms traditional cellular network operations by introducing programmable Radio Intelligence Controller (RIC) capabilities that enable intelligent network optimisation through xApps and rApps \cite{d2022dapps}. The RIC leverages key performance monitoring (KPM) data transmitted over the E2 interface to provide real-time network insights, enabling dynamic Radio Unit (RU) management decisions that can significantly improve energy efficiency (EE) while maintaining service quality \cite{hoffmann2023open}. This programmable approach represents a paradigm shift from static network configurations to adaptive, intelligence-driven resource management that can respond to varying traffic conditions and operational requirements.

\begin{figure}
\centering
\includegraphics[width=\columnwidth]{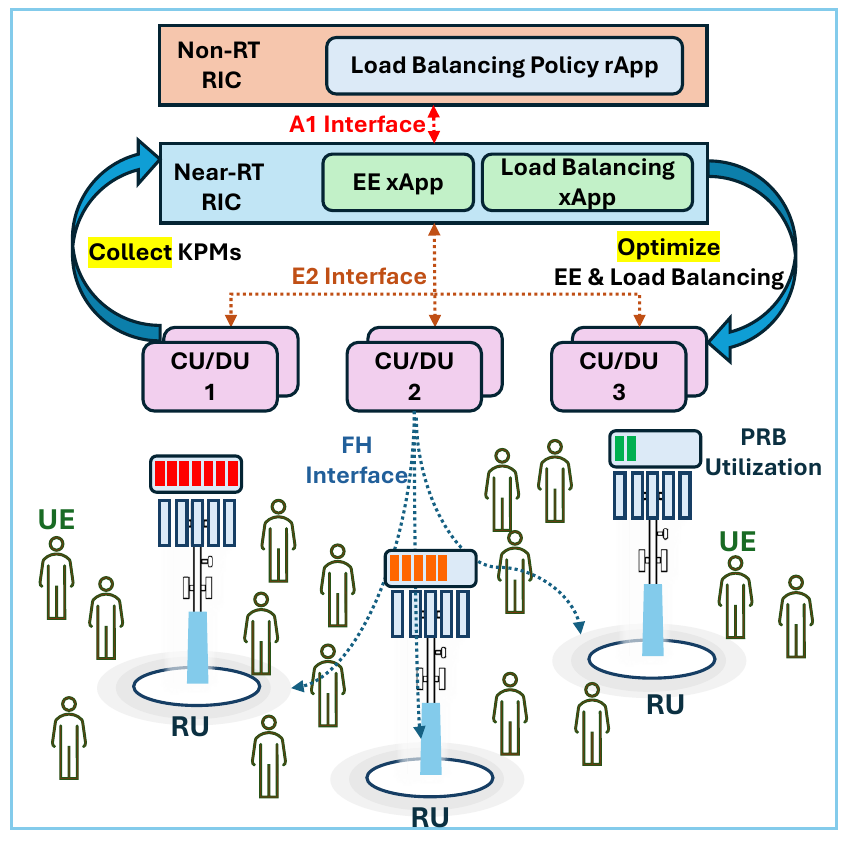}
\caption{System Model}
\label{fig:sys_env}
\end{figure}

The inherent flexibility of O-RAN architecture enables xApps to implement dynamic RU sleeping strategies, where underutilised RUs can be temporarily deactivated to reduce power consumption during low-traffic periods. Several existing artificial intelligence (AI) and machine learning (ML) based approaches have demonstrated the potential for substantial energy savings through such dynamic sleeping mechanisms \cite{qazzaz2025oran, vallero2025threshold}. These studies typically report energy reductions of 30-45\% by intelligently switching off RUs based on traffic load predictions and spatial coverage requirements.

Nevertheless, most existing approaches rely predominantly upon user equipment (UE) offloading strategies to create sleeping opportunities for target RUs. While this approach effectively reduces the number of active RUs, it inherently creates significant load imbalance across the remaining active network infrastructure \cite{el2022energy, post2021self}. The fundamental issue arises when UEs originally served by a sleeping RU are transferred to neighbouring active RUs, resulting in uneven traffic distribution where some RUs become heavily loaded while others remain underutilised. Such load imbalance may severely impact the throughput performance of offloaded UEs as they might be allocated a substantially lower number of physical resource blocks (PRBs) due to increased competition for limited radio resources on the serving RUs.

The challenge becomes even more complex when attempting to maintain equivalent PRB allocation for all UEs while simultaneously addressing energy efficiency objectives at the network level. Traditional energy-first optimisation strategies typically prioritise power savings without adequately considering the load distribution consequences, often resulting in performance bottlenecks that negate the benefits of energy reduction \cite{ichimescu2024energy}. Conversely, load balancing approaches focus on uniform traffic distribution but operate under fixed RU configurations, thereby missing opportunities for intelligent energy management through dynamic RU switching.

Current machine learning applications in RAN optimisation, including reinforcement learning~\cite{sun2024deep}, deep neural networks~\cite{bordin2025design}, and ensemble methods~\cite{ntassah2024user}, predominantly address single-objective optimisation problems. These approaches either focus on energy efficiency maximisation or performance optimisation in isolation, failing to capture the intricate interdependencies between RU configuration decisions, traffic distribution patterns, and network performance metrics. Moreover, most existing ML frameworks operate reactively, responding to load imbalances after RU switching decisions have been implemented, rather than predictively evaluating the load distribution implications of proposed energy optimisation actions.

To that end, in this article, we present a comprehensive ML-based framework for joint optimisation of load balancing and energy efficiency for O-RAN deployments. We formulate the problem as a multi-class classification system that predictively evaluates potential RU configurations before implementing energy efficiency optimisations, enabling proactive selection of RU switching strategies that maintain acceptable load distribution. Our approach maps network conditions to three distinct load balance categories (Well Balanced, Moderately Balanced, Imbalanced) using engineered features extracted from real-time KPM data available through the E2 interface.

Future research could explore the integration of deep reinforcement learning techniques to enable adaptive threshold policy selection that continuously learns optimal Conservative/Moderate/Aggressive strategies through real-time interaction with dynamic network environments. This approach can automatically adjust classification boundaries based on historical performance patterns, seasonal traffic variations, and emerging network conditions, eliminating the need for manual reconfiguration.

The remainder of this paper is organised as follows. Section~\ref{sys_model} presents the System Model and Methodology, which includes three subsections: Section~\ref{net_model} detailing the O-RAN network model and mathematical formulation; Section~\ref{load} describes the multi-class classification approach and threshold configurations; and Model Deployment outlines the rApp/xApp architecture and baseline validation strategies are presented in Section~\ref{deploy}. Section~\ref{sec:results} presents the Experimental Setup and Results, analysing the performance of the proposed ML framework in terms of load balance prediction accuracy, network performance improvements, and comparison against baseline strategies. Finally, Section~\ref{concl} concludes the paper, summarising key findings of this research.

\section{System Model and Methodology}\label{sys_model}

\subsection{Network Model}\label{net_model}

We consider an O-RAN deployment consisting of $N$ configurable RUs and $U$ UEs distributed across the coverage area as shown in Fig.~\ref{fig:sys_env}. Each RU $i \in \{1, 2, \ldots, N\}$ operates in one of two binary states: active ($s_i = 1$) or inactive ($s_i = 0$), forming the network configuration vector $\mathbf{S}(t) = [s_1(t), s_2(t), \ldots, s_N(t)]$ at time $t$. The number of active RUs is $N_{active}(t) = \sum_{i=1}^{N} s_i(t)$.

The traffic load per RU is characterised by the PRB utilisation vector $\mathbf{P}(t) = [p_1(t), p_2(t), \ldots, p_N(t)]$, where $p_i(t)$ denotes the downlink PRB usage percentage of RU $i$ at time $t$. Inactive RUs contribute zero load ($p_i(t) = 0$ when $s_i(t) = 0$). The aggregate network performance is measured through QoS metrics including throughput, latency, and overall QoS score, denoted as $Q(t) = f(\mathbf{S}(t), \mathbf{P}(t), \mathbf{U}(t))$, where $\mathbf{U}(t)$ represents the UE distribution and traffic demand vector.

Our system addresses intelligent RU configuration selection for energy efficiency while maintaining load balance. Rather than operating reactively after RU switching decisions, our ML model predictively evaluates potential RU configurations before implementation, enabling operators to achieve energy savings while avoiding configurations that cause load imbalances and performance bottlenecks. The system enables the proactive selection of RU configurations that achieve the desired energy savings while maintaining the predicted load balance quality.

\subsection{Load Balancing Framework}\label{load}

To quantify traffic distribution balance across active RUs, we compute three key metrics that collectively offer a robust view of balance quality:

\textbf{Coefficient of Variation (CV):} Measures relative variation of load distribution across active RUs:
\begin{equation}
CV_{DL}(t) = \frac{\sigma(\mathbf{P}_{active}(t))}{\mu(\mathbf{P}_{active}(t))}
\end{equation}
where $\mathbf{P}_{active}(t)$ contains only active RU loads, $\sigma$ is the standard deviation, and $\mu$ is the mean. Lower CV values indicate better load balance.

\textbf{Jain's Fairness Index:} Evaluates load uniformity across active RUs:
\begin{equation}
J_{DL}(t) = \frac{(\sum_{i \in A} p_i(t))^2}{|A| \cdot \sum_{i \in A} p_i^2(t)}
\end{equation}
where $A$ is the set of active RUs and $|A|$ is the cardinality. Values closer to 1 indicate better fairness.

\textbf{Load Imbalance Factor (LIF):} Quantifies the extent to which the most heavily loaded RU exceeds the average load:
\begin{equation}
LIF(t) = \frac{p_{max}(t)}{\mu(\mathbf{P}_{active}(t))} - 1
\end{equation}
where $p_{max}(t)$ is the maximum load among active RUs. A value of 0 indicates perfect balance, while larger values indicate greater imbalance concentrated in the most loaded RU.

We formulate load balancing as a multi-class classification problem mapping network states to three categories: \textbf{Well Balanced} (minimal load variation, optimal utilisation), \textbf{Moderately Balanced} (acceptable distribution with optimisation potential), and \textbf{Imbalanced} (significant variation causing bottlenecks). This classification enables the prediction of whether proposed RU configurations will achieve balanced traffic distribution before implementation.

The classification decision is made by comparing the computed $CV_{DL}$, $J_{DL}$, and $LIF$ values against predefined thresholds. For each threshold configuration $\tau \in {\text{Conservative}, \text{Moderate}, \text{Aggressive}}$, we define classification boundaries using five threshold parameters: $\alpha_\tau$ (CV upper bound for Well Balanced), $\beta_\tau$ (Jain's index lower bound for Well Balanced), $\gamma_\tau$ (LIF upper bound for Well Balanced), $\delta_\tau$ (CV upper bound for Moderately Balanced), and $\epsilon_\tau$ (Jain's index lower bound for Moderately Balanced):

\begin{equation}
\text{Category}_\tau(t) = \begin{cases}
\text{Well Balanced} & \text{if } CV_{DL} \leq \alpha_\tau \land \\
& J_{DL} \geq \beta_\tau \land LIF \leq \gamma_\tau \\
\text{Mod. Balanced} & \text{if } \alpha_\tau < CV_{DL} \leq \delta_\tau \\
& \text{or } \epsilon_\tau \leq J_{DL} < \beta_\tau \\
\text{Imbalanced} & \text{otherwise}
\end{cases}
\end{equation}

To support different operational strategies, we define three threshold configurations with specific parameter values:
\begin{align}
\text{Conservative}: &\quad \alpha = 0.3, \beta = 0.8, \gamma = 1.0, \delta = 0.5, \epsilon = 0.7 \\
\text{Moderate}: &\quad \alpha = 0.5, \beta = 0.7, \gamma = 1.5, \delta = 0.7, \epsilon = 0.6 \\
\text{Aggressive}: &\quad \alpha = 0.7, \beta = 0.6, \gamma = 2.0, \delta = 0.9, \epsilon = 0.5
\end{align}

The threshold design reflects operational priorities: \textbf{Conservative} uses strict criteria ($\alpha = 0.3$, $\beta = 0.8$, $\gamma = 1.0$) for critical infrastructure requiring perfect balance, \textbf{Moderate} provides balanced requirements ($\alpha = 0.5$, $\beta = 0.7$, $\gamma = 1.5$) for standard operations, and \textbf{Aggressive} employs relaxed criteria ($\alpha = 0.7$, $\beta = 0.6$, $\gamma = 2.0$) prioritizing energy savings over perfect balance. This progressive design enables context-aware evaluation based on network criticality and energy efficiency priorities.

\begin{algorithm}[t]
\caption{Real-time RU Configuration Selection for Energy Optimisation}
\label{alg:ru_configuration}
\begin{algorithmic}[1]
\STATE \textbf{Input:} Current network state, threshold policy $\tau$, trained model $M_{\tau}$
\STATE \textbf{Output:} Optimal RU configuration $\mathbf{S}_{optimal}$
\STATE Initialize candidate configurations $\mathcal{C} = \emptyset$
\FOR{each potential RU deactivation scenario}
    \STATE Generate candidate configuration $\mathbf{S}_{candidate}$
    \STATE Predict load balance: $\text{category} = M_{\tau}(\mathbf{S}_{candidate})$
    \IF{category $\in$ \{Well Balanced, Moderately Balanced\}}
        \STATE Add $(\mathbf{S}_{candidate}, \text{energy\_savings})$ to $\mathcal{C}$
    \ENDIF
\ENDFOR
\IF{$\mathcal{C} \neq \emptyset$}
    \STATE Select configuration with maximum energy savings from $\mathcal{C}$
\ELSE
    \STATE Maintain current configuration (no acceptable energy-saving option)
\ENDIF
\STATE Return $\mathbf{S}_{optimal}$
\end{algorithmic}
\end{algorithm}

We evaluate four supervised ML models: \textbf{Random Forest} (100 trees, depth = 10), \textbf{XGBoost} (gradient boosting), \textbf{Logistic Regression} (L2 regularisation), and \textbf{Neural Network} (hidden layers [128, 64]). Random Forest serves as our primary model due to its superior performance and interpretability. Performance is evaluated using the F1-macro score, which averages F1-scores across all categories to ensure balanced evaluation despite class imbalance.

\subsection{Model Deployment}\label{deploy}

The system operates within the O-RAN architecture through two complementary components designed for different operational timescales as presented in Figure~\ref{fig:sys_env}. An \textbf{rApp} deployed in the Non-RT RIC manages strategic policies and threshold selection based on long-term network analysis (operating on $>1$ second timescales. In contrast, an \textbf{xApp} in the Near-RT RIC executes real-time ML-driven RU configuration decisions (operating within 10ms-1s timescales) \cite{d2022dapps}.

The deployment workflow operates as follows: (1) the rApp analyzes network context including location type, time of day, and traffic patterns to select appropriate threshold policies—Conservative for critical areas requiring maximum reliability, Moderate for standard operations balancing performance and efficiency, and Aggressive for scenarios prioritizing maximum energy savings; (2) the xApp receives the current threshold policy and continuously monitors network state through real-time telemetry data collection; (3) when energy optimization opportunities are identified, the xApp generates candidate RU configurations representing different energy-saving scenarios; (4) for each candidate configuration, the ML model predicts the resulting load balance quality; (5) the model then selects the configuration that achieves desired energy savings while maintaining acceptable load distribution according to the current threshold policy. Algorithm~\ref{alg:ru_configuration} formalises this decision-making process, showing how the xApp systematically evaluates candidate RU configurations and selects the optimal energy-saving solution that maintains predicted load balance quality.

To validate ML performance against current industry practices, we implement six baseline strategies representing different operational approaches: (i) \textbf{Random Baseline} assigns load balance categories with the same probability distribution as the training data, simulating unguided decision-making; (ii) \textbf{Energy-First Strategy} always predicts "Well Balanced" to justify minimizing active RUs for maximum power savings, regardless of actual load conditions; (iii) \textbf{Conservative Strategy} always predicts "Imbalanced" to maintain all RUs active, prioritizing service reliability over energy efficiency; (iv) \textbf{Majority Class Baseline} consistently predicts the most frequent category from training data, representing a naive statistical approach; (v) \textbf{RU-Count Rule} applies a simple heuristic where load balance category depends solely on the number of active RUs; (vi) \textbf{Load-Based Rule} uses threshold-based decisions based on average PRB utilization levels across active RUs.

\section{Experimental Setup and Results}\label{sec:results}

\subsection{Experimental Setup}

Our experimental evaluation utilises a comprehensive dataset of 4.26 million network snapshots generated through O-RAN digital twin simulations. The simulation environment models realistic O-RAN deployments with three distinct RU configuration scenarios: 4, 5, and 6 RUs serving 30 UEs with uniform spatial distribution and 30/70 uplink/downlink traffic split, representing typical cellular network patterns. The dataset encompasses critical performance metrics, including: PRB utilisation indicating spectrum usage efficiency, QoS Score (composite measure combining throughput, latency, and service quality indicators), UE attachment and mobility patterns, power consumption measurements across RU configurations, and throughput measurements for both uplink and downlink traffic.

From these raw measurements, we extract 47 engineered features organized into five categories: load distribution features (PRB utilization statistics including mean, standard deviation, percentiles), resource utilization features (efficiency metrics and utilization ratios), connection pattern features (UE distribution characteristics and RRC connection statistics), traffic characteristics features (asymmetry measures and temporal variations), and performance indicators (QoS-related features including power efficiency and throughput per RU)to evaluate potential RU configurations before implementation predictivelylex relationships between network state and load balance quality.

All models are trained for deployment within the O-RAN architecture, where threshold policies are selected by an rApp in the Non-RT RIC based on operational context (location type, time of day, traffic patterns), and real-time load balance predictions are executed by an xApp in the Near-RT RIC to guide intelligent RU configuration decisions. The system is designed to predictively evaluate potential RU configurations before implementation, enabling proactive energy savings while avoiding configurations that cause load imbalances.

Model implementation uses Python scikit-learn with stratified 5-fold cross-validation and 70\%/15\%/15\% train/validation/test splits to ensure robust evaluation across diverse network conditions. The ML models are configured with optimized hyperparameters: Random Forest with $n_{trees} = 100$ and maximum depth = 10, XGBoost with learning rate = 0.1 and maximum depth = 6, Logistic Regression with L2 regularization ($C = 1.0$) and balanced class weights, and Neural Network with hidden layers [128, 64], ReLU activation, and Adam optimizer with learning rate = 0.001.

\subsection{Results and Discussions}

We evaluate models' prediction performance using F1-macro score, which averages F1-scores across all categories to ensure equal consideration of Imbalanced, Moderately Balanced, and Well Balanced predictions regardless of their frequency in the dataset. 

\begin{figure}
\centering
\includegraphics[width=\columnwidth]{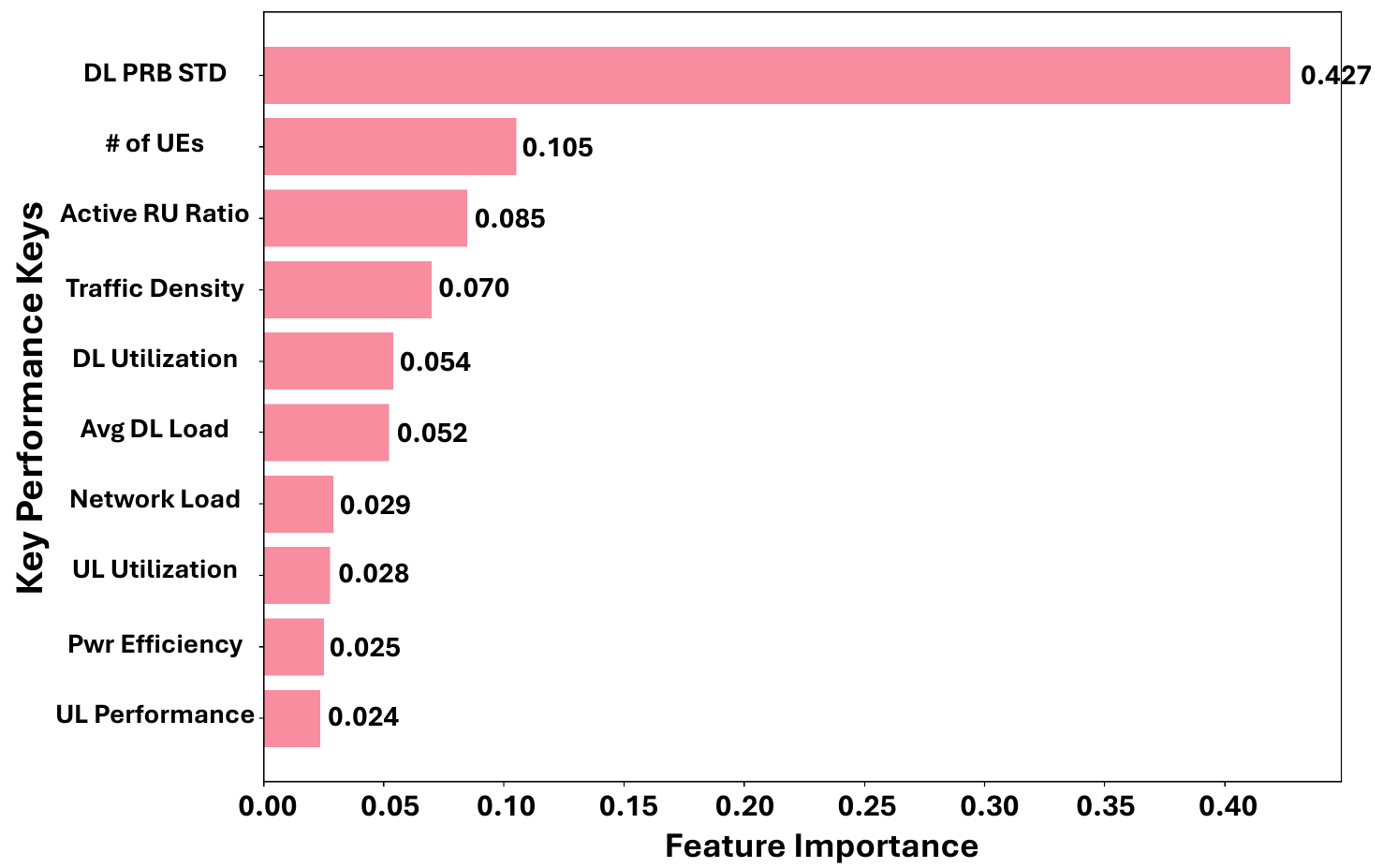}
\caption{Top 10 feature importance for RU Configuration Decisions}
\label{fig:feature_importance}
\end{figure}

Feature importance analysis reveals that load variation metrics dominate RU configuration decisions, with downlink PRB standard deviation (DL PRB STD) contributing 42.7\% of model predictions. Fig.~\ref{fig:feature_importance} illustrates the top 10 most influential features for RU configuration decisions, validating our belief that traffic distribution variation, over absolute load levels, is the critical factor for intelligent RU switching decisions. Secondary important features include UE connection patterns (\# of UEs, 10.5\%) and active RU efficiency ratios (8.5\%), demonstrating the importance of both load variation and resource allocation patterns in O-RAN optimisation.

\begin{table}
\centering
\caption{ML Model Performance Comparison}
\label{tab:ml_performance}
\resizebox{\columnwidth}{!}{%
\begin{tabular}{|l|c|c|c|c|}
\hline
\textbf{Model} & \textbf{Threshold} & \textbf{Accuracy} & \textbf{F1-Macro} & \textbf{CV Score} \\
\hline
\multirow{3}{*}{Random Forest} & Conservative & 1.000 & 0.880 & 0.852 ± 0.038 \\
& Moderate & 0.996 & 0.966 & 0.957 ± 0.004 \\
& Aggressive & 0.994 & 0.983 & 0.981 ± 0.002 \\
\hline
\multirow{3}{*}{XGBoost} & Conservative & 0.999 & 0.647 & 0.617 ± 0.044 \\
& Moderate & 0.995 & 0.894 & 0.891 ± 0.005 \\
& Aggressive & 0.994 & 0.977 & 0.977 ± 0.001 \\
\hline
\multirow{3}{*}{Logistic Regression} & Conservative & 0.999 & 0.362 & 0.356 ± 0.007 \\
& Moderate & 0.992 & 0.629 & 0.648 ± 0.009 \\
& Aggressive & 0.990 & 0.950 & 0.942 ± 0.005 \\
\hline
\multirow{3}{*}{Neural Network} & Conservative & 0.999 & 0.338 & 0.532 ± 0.040 \\
& Moderate & 0.995 & 0.921 & 0.905 ± 0.045 \\
& Aggressive & 0.989 & 0.967 & 0.941 ± 0.055 \\
\hline
\end{tabular}%
}
\end{table}

\begin{figure}
\centering
\includegraphics[width=\columnwidth]{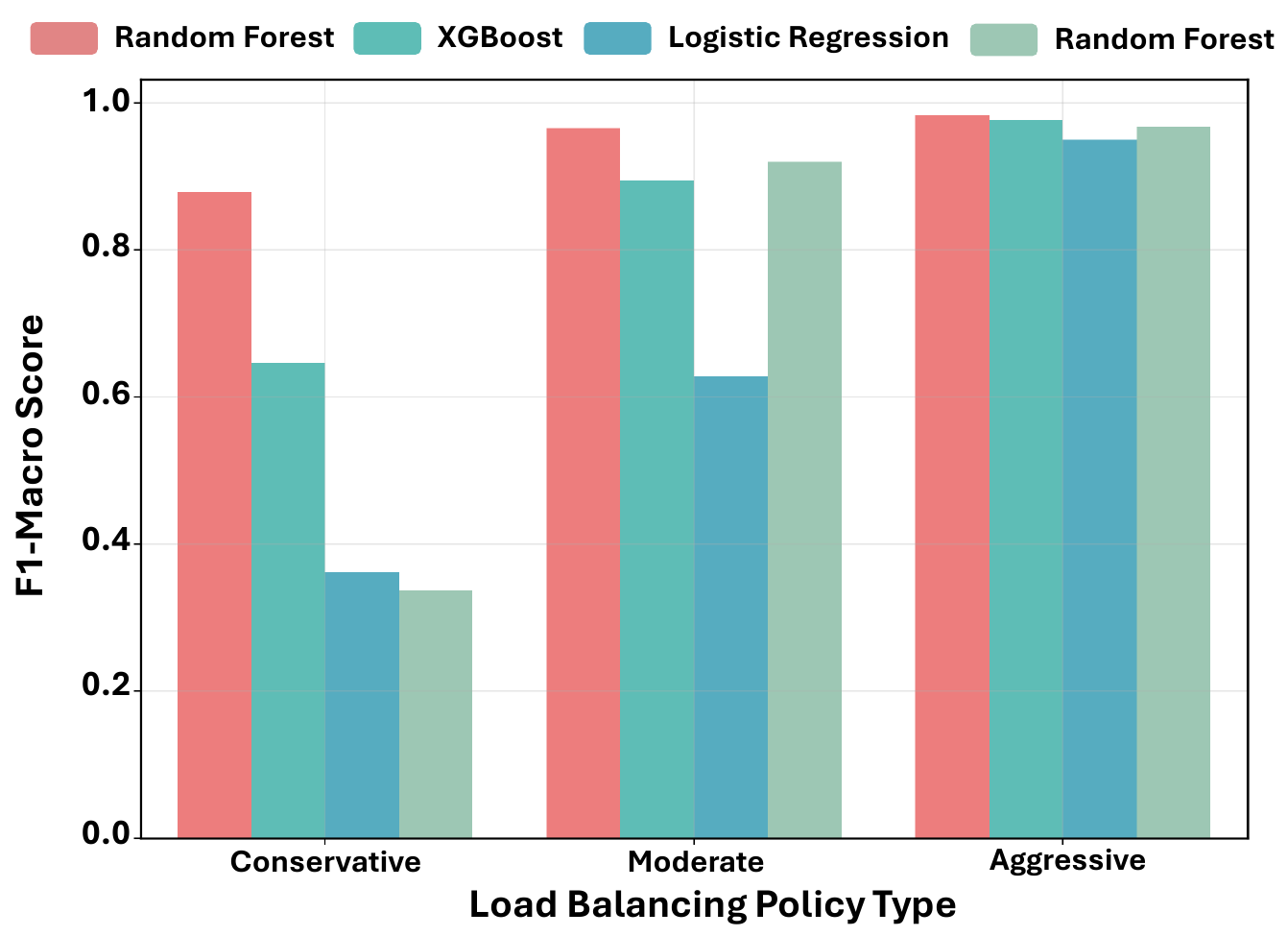}
\caption{F1-macro performance comparison across ML models and threshold configurations for load balance prediction quality}
\label{fig:f1_comparison}
\end{figure}

Fig.~\ref{fig:f1_comparison} presents the comprehensive performance comparison across all ML models and threshold configurations for predicting the load balance quality of proposed RU configurations. Random Forest consistently achieves superior decision-making accuracy, with F1-macro scores of 88.0\%, 96.6\%, and 98.3\% for Conservative, Moderate, and Aggressive thresholds, respectively. The Moderate threshold provides optimal balance for standard network operations, achieving 96.6\% F1-macro with excellent cross-validation stability (95.7\% ± 0.4\%), enabling reliable RU configuration selection for typical deployment scenarios.

XGBoost demonstrates competitive performance in Aggressive configurations (97.7\% F1-macro), suitable for maximum energy savings scenarios, while Logistic Regression and Neural Network show progressively lower performance across all operational contexts. The detailed performance metrics are summarised in Table~\ref{tab:ml_performance}, which indicates that Random Forest maintains the most consistent performance across all threshold configurations with superior cross-validation scores.

\begin{figure}
\centering
\includegraphics[width=0.925\columnwidth]{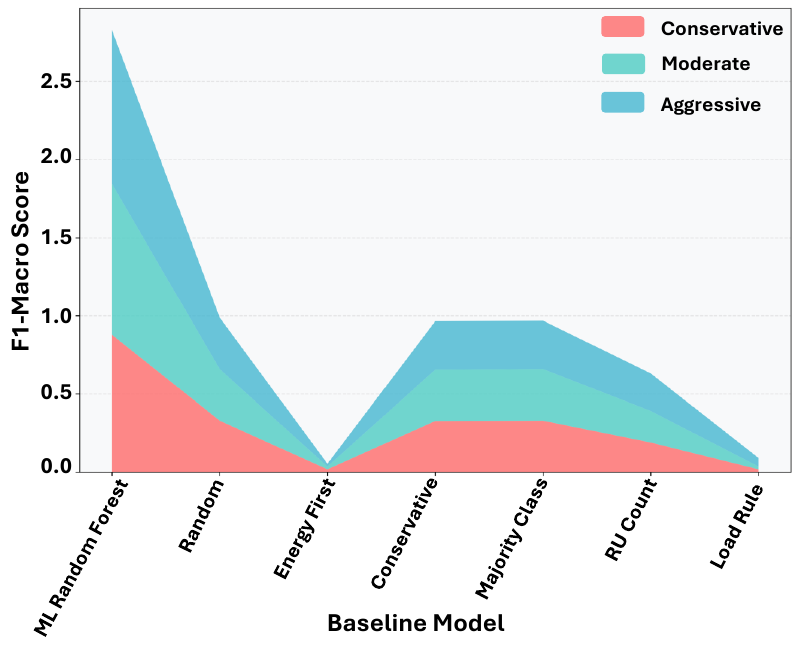}
\caption{Random Forest model vs baseline strategy comparison for RU configuration decision quality}
\label{fig:baseline_f1_comparison}
\end{figure}

The dramatic superiority of ML-driven decision making over traditional RU switching strategies is demonstrated in Fig.~\ref{fig:baseline_f1_comparison}. Our Random Forest model achieves substantial improvements in predicting optimal RU configurations: 191\% over Random selection strategies, 194\% over Conservative approaches that maintain all RUs active, and remarkable gains exceeding 5000\% over Energy-First strategies that ignore load distribution consequences. Even sophisticated RU-Count heuristics achieve only 20.2\% F1-macro, demonstrating the critical need for intelligent load balance prediction in dynamic RU switching decisions.

\begin{figure}
\centering
\includegraphics[width=\columnwidth]{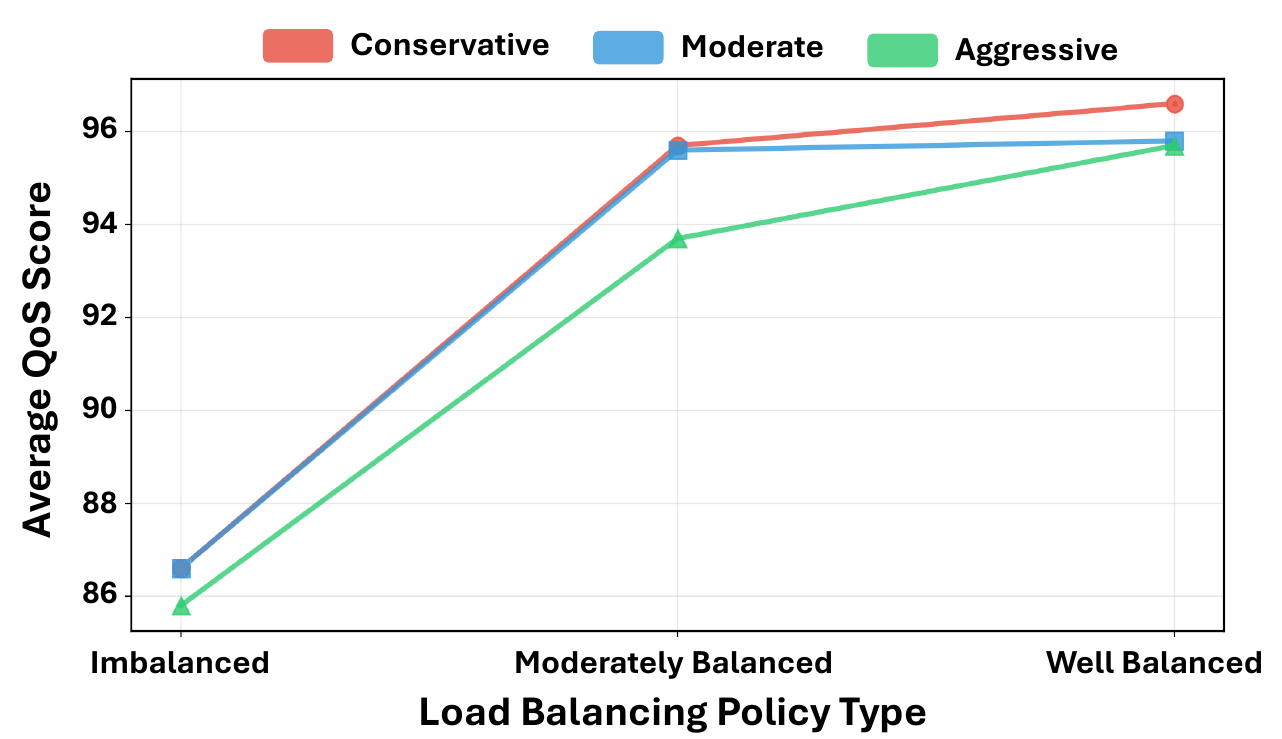}
\caption{QoS performance by predicted load balance category}
\label{fig:network_performance_impact}
\end{figure}

Network performance analysis validates the practical benefits of ML-guided RU configuration selection. Fig.~\ref{fig:network_performance_impact} demonstrates QoS improvements achievable through accurate load balance prediction, showing 10.6-11.6\% gains when ML models correctly identify Well Balanced configurations versus Imbalanced states. The Moderate threshold configuration shows QoS progression from 86.6 (Imbalanced) to 95.8 (Well Balanced), proving that accurate load balance prediction directly translates to network performance improvements in operational deployments.

Table~\ref{tab:network_impact} summarises the quantitative network performance implications across all load balance categories, revealing the practical benefits achievable through optimal load balance prediction. The analysis demonstrates that transitioning from Imbalanced to Well Balanced categories achieves up to 10.6\% QoS improvement and 71.1\% better load distribution, while requiring only moderate power increases (22\%) that are justified by the substantial performance gains.

\begin{table}
\centering
\caption{Network Performance Impact by Load Balance Category (Moderate Threshold)}
\label{tab:network_impact}
\resizebox{\columnwidth}{!}{%
\begin{tabular}{|l|c|c|c|}
\hline
\textbf{Category} & \textbf{QoS Score} & \textbf{Load Balance CV} & \textbf{Power (W)} \\
\hline
Imbalanced & 86.62 & 1.458 & 23.16 \\
Mod. Balanced & 95.63 & 0.656 & 26.38 \\
Well Balanced & 95.84 & 0.422 & 28.28 \\
\hline
\textbf{Improvement Potential} & \textbf{+10.6\%} & \textbf{-71.1\%} & \textbf{Energy Trade-off} \\
\hline
\end{tabular}%
}
\end{table}

\begin{figure}
\centering
\includegraphics[width=0.9\columnwidth]{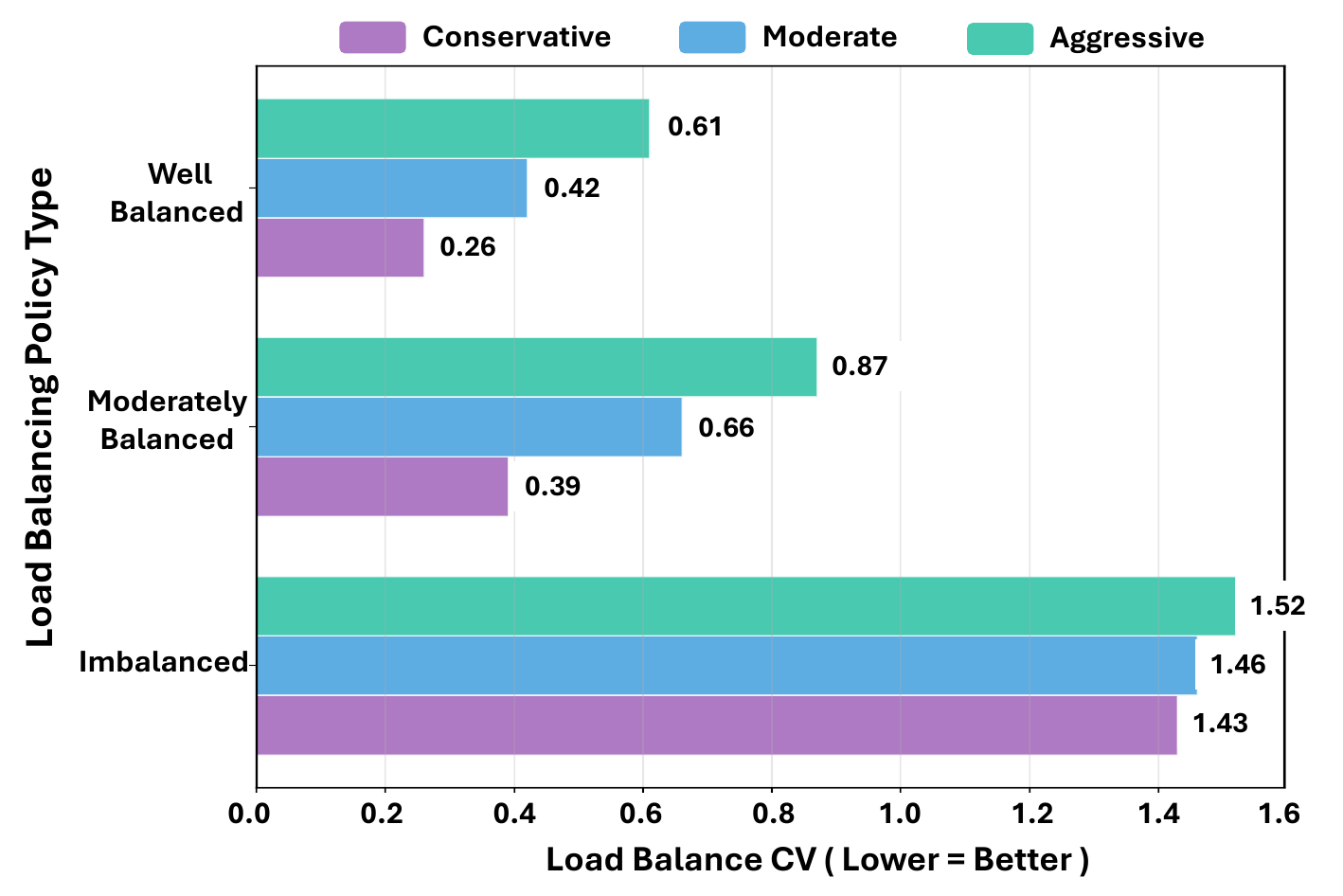}
\caption{Load distribution quality by predicted category}
\label{fig:load_balance_improvement}
\end{figure}

The load distribution benefits of ML-guided RU switching are illustrated in Fig.~\ref{fig:load_balance_improvement} through coefficient of variation analysis. ML-predicted Well Balanced configurations achieve superior traffic distribution uniformity (CV: 0.26-0.42) compared to Imbalanced states (CV: 1.43-1.52), representing up to 82\% improvement in load distribution quality. This validates that our ML framework enables operators to select RU configurations that prevent traffic bottlenecks while achieving energy savings, addressing the fundamental challenge of dynamic RU switching optimisation.

\section{Conclusions}\label{concl}

This paper presents a comprehensive machine learning framework for joint optimisation of load balancing and energy efficiency in O-RAN deployments. The proposed multi-class classification system enables predictive evaluation of potential RU configurations before implementation, shifting from reactive to proactive load balancing strategies.

Experimental evaluation using 4.26 million real network measurements demonstrates substantial superiority over traditional approaches. The Random Forest model achieves 98.3\% F1-macro score, representing 195\% improvement over baseline strategies, while enabling up to 11.6\% QoS improvement and 82\% better load distribution. The multi-threshold approach provides operational flexibility across diverse deployment scenarios, from critical infrastructure to energy-prioritised rural networks.

\section*{Acknowledgment}

This research was funded by EP/X040518/1 EPSRC CHEDDAR, UKRI Grant EP/X039161/1, and MSCA Horizon EU Grant 101086218 and partially by the UKRI Funding Service under Award UKRI851.

\bibliographystyle{IEEEtran}
\bibliography{IEEEabrv,Bibliography}

\end{document}